\documentclass[sigconf,authorversion,nonacm]{acmart}

\usepackage{natbib} 
\usepackage{multirow} 
\usepackage{tabularx} 
\usepackage[ruled,linesnumbered]{algorithm2e} 
\usepackage{xspace} 
\usepackage{pgfplots}\pgfplotsset{compat=1.9} 
\usepackage{balance} 
\usepackage{soul} 

\usepackage{xcolor}

\newcommand{\new}[1]{#1}
\newcommand{\neww}[1]{#1}

\newcolumntype{L}[1]{>{\raggedright\let\newline\\\arraybackslash\hspace{0pt}}m{#1}}
\newcolumntype{C}[1]{>{\centering\let\newline\\\arraybackslash\hspace{0pt}}m{#1}}
\newcolumntype{R}[1]{>{\raggedleft\let\newline\\\arraybackslash\hspace{0pt}}m{#1}}

\theoremstyle{definition}
\newtheorem{definition}{Notation}[section]

\newcommand{\textss}[1]{\textsuperscript{#1}}
\newcommand{\Glob}{SED\xspace} 
\newcommand{\tempo}{TED\xspace} 
\newcommand{\Tempo}{TED\xspace} 
\newcommand{\static}{static\xspace}
\newcommand{\Static}{Static\xspace}

\DeclareMathOperator*{\argmin}{arg\,min}

\begin{document}

\title{Event-Driven Query Expansion}

\author{Guy D. Rosin}
\affiliation{
  \institution{Technion}
  \city{Haifa}
  \country{Israel}
}
\email{guyrosin@cs.technion.ac.il}
\author{Ido Guy}
\affiliation{
  \institution{eBay Research}
  \city{Netanya}
  \country{Israel}
}
\email{idoguy@acm.org}
\author{Kira Radinsky}
\affiliation{
  \institution{Technion}
  \city{Haifa}
  \country{Israel}
}
\email{kirar@cs.technion.ac.il}


\begin{abstract}

A significant number of event-related queries are issued in Web search.
In this paper, we seek to improve retrieval performance by leveraging events and specifically target the classic task of query expansion. 
We propose a method to expand an event-related query by first detecting the events related to it.
Then, we derive the candidates for expansion as terms semantically related to both the query and the events.
To identify the candidates, we utilize a novel mechanism to simultaneously embed words and events in the same vector space.
We show that our proposed method of leveraging events improves query expansion performance significantly compared with state-of-the-art methods on various newswire TREC datasets.

\end{abstract}



\maketitle

\section{Introduction}
\label{sec:intro}

One of the common practices in Information Retrieval (IR) is the enrichment of search queries with additional terms~\citep{carpineto2012qesurvey, azad2019qesurvey}.
This approach is called query expansion (QE) and is intended to narrow the gap between the query and the corpus language, to increase retrieval's recall. In this work, we focus on the task of QE for event-related queries.

A significant number of event-related queries are issued in Web search. Based on our analysis over an AOL query log~\citep{pass2006aoldata}, $36\%$ of Web search queries 
are related to events (evaluation method described in Section~\ref{sec:event_related_query_classification}).
While several studies have focused on detecting such queries~\citep{ghoreishi2013predicting,kanhabua2015learning,zhang2018automatic}, no attempt has been made to leverage event representations for QE.

While some event-related queries explicitly include a temporal expression~\citep{nunes2008use} (e.g., ``2020 pandemic'' or ``US election 2016''), most are implicitly associated with a particular time period~\citep{kanhabua2010determining} (e.g., ``Black Monday'' or ``Tiananmen Square protesters'').
In addition, queries can be relevant to either a single event (``Schengen agreement'') or multiple events (``computer viruses'').

We suggest adding an intermediate step in query expansion of identifying related events. Our hypothesis is that it can help determine the right time span and context for search results, and thus help guide the query expansion.
\new{Detecting the right events is a challenging task for several reasons.
First, queries are usually very short and may contain ambiguous terms. For example, given the query ``Bank Failures'', our method detected the event ``2000 Camp David Summit'', which discussed the West Bank territory and is considered a failure.\footnote{\url{https://en.wikipedia.org/wiki/2000_Camp_David_Summit}}
Second, events can be relevant to entities in the query, but not to the query intent. As an example, given the query ``Greenpeace'', our method detected ``Sinking of the Rainbow Warrior'' (a bombing of Greenpeace's flagship). This event is indeed relevant to Greenpeace, but it does not fit the query intent and thus it does not provide a relevant context for query expansion.}

\begin{figure}
\centering
\includegraphics[width=\linewidth]{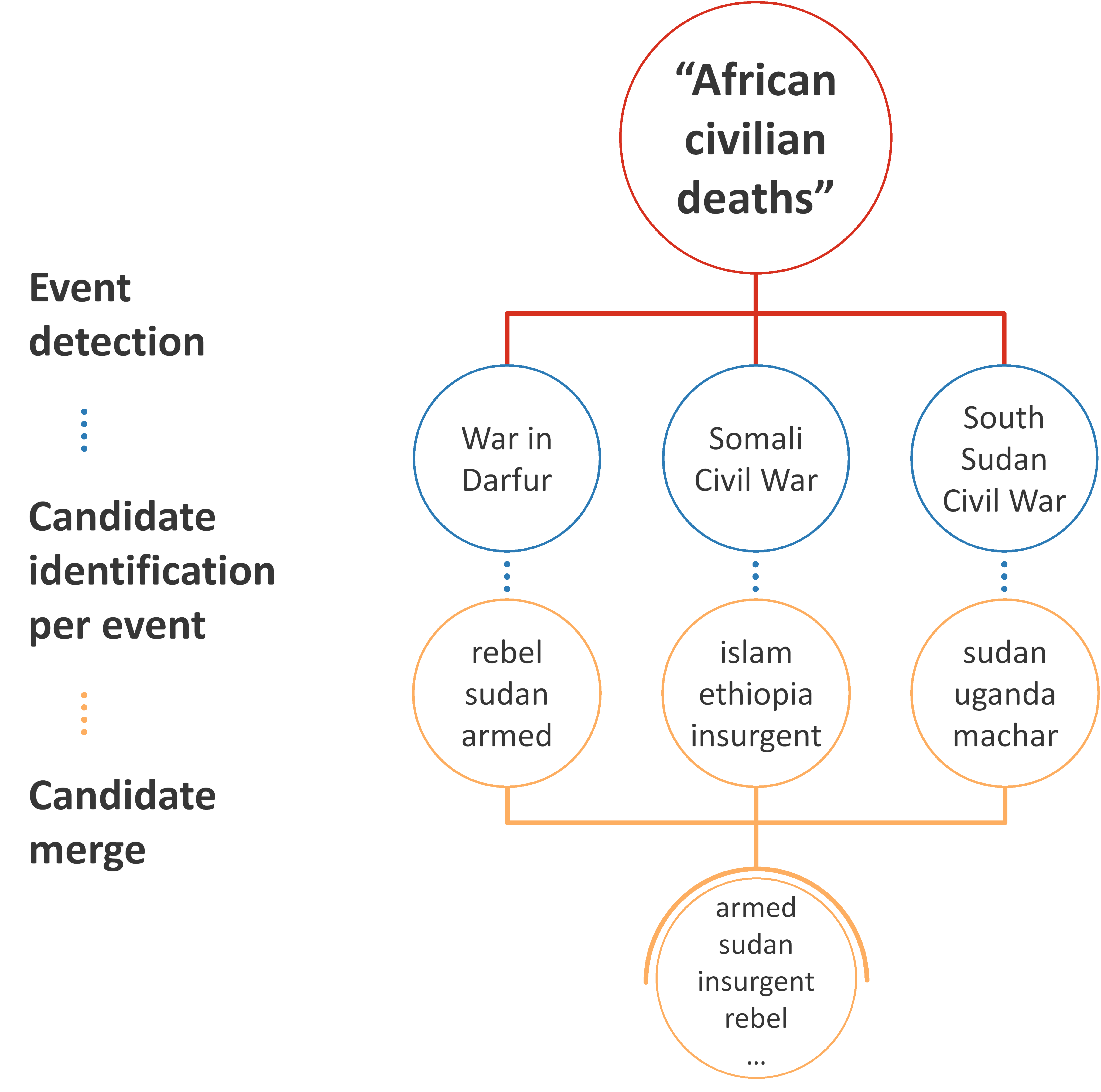}
\caption{\label{fig:pipeline}Pipeline of event-driven query expansion, with the example query of ``African civilian deaths''.}
\end{figure}

We propose a method to expand an event-related query by first detecting the events related to it. Then, based on the detected events, we derive the candidates for expansion as terms semantically related to both the query and the events. 
\new{Figure~\ref{fig:pipeline} shows an example: given the query ``African civilian deaths'', our method detects the following events: \textit{War in Darfur}, \textit{Somali Civil War}, and \textit{South Sudanese Civil War}. Then, it suggests these top expansions: `armed', `sudan', `insurgent', `rebel', and `uganda'. These are all highly relevant events and expansions for the query.
On the other hand, a baseline method \cite{imani2019deep} produced as top expansions names of African nationalities: `sudanese', `rhodesian', and `ugandan'.}

We present a novel technique for embedding both words and events in the same vector space (Section~\ref{sec:event_projection}). 
\new{Previous work has often used large static corpora, such as Wikipedia, to represent words~\citep{gabrilovich2007computing,word2vec,yamada2016joint}. Similarly, events can be represented by the embedding of their corresponding article in Wikipedia~\citep{gabrilovich2007computing,das2017estimating}.}
However, events occur at a specific time, hence their semantics is true only for that time~\citep{rosin2019generating}.
Static representations, such as Wikipedia, capture a description of the event in perspective to today, thus losing their temporal representation.
For example, a major part of World War II's Wikipedia entry\footnote{\url{https://en.wikipedia.org/wiki/World_War_II}} is dedicated to periods before or after the war itself (i.e., pre-war events, its aftermath and impact).
We therefore conjecture that the use of temporal embeddings would enable to better capture the interaction between the event and the QE candidates.
We suggest to project static event embeddings onto a temporal corpus (38 years of the \textit{New York Times}).
This allows tracking words and events dynamics over time (Section~\ref{sec:temporal_embeddings}), to identify potential candidates for expansion.
Intuitively, we identify words whose semantics is influenced by the event and hypothesize that they serve as better expansion candidates for the query. 
\neww{Since the proposed projection method is applied as a preprocessing step, our QE method is suitable for a real-time search scenario.}
Our empirical results over the global analysis approach for automatic query expansion~\citep{azad2019qesurvey}, show high performance gains compared with state-of-the-art query expansion algorithms~\citep{padaki2020rethinking,imani2019deep,kuzi2016query,roy2016qe}.

The contributions of this paper are threefold:
First, we propose an approach for query expansion by adding an intermediate step of identifying events related to the query (Section~\ref{sec:event_detection}).
Those, in turn, are used to identify more precise query expansion candidates (Section~\ref{sec:event_driven_expansion}).
Second, we present a novel mechanism to simultaneously embed words and events in the same space (Section~\ref{sec:temporal_embeddings}).
The embeddings are used to identify expansion candidates whose semantics was influenced by the events related to the query.
Third, we show that our proposed method of leveraging events improves QE performance significantly (Section~\ref{sec:results}).
We also publish our code and data.\footnote{\url{https://github.com/guyrosin/event_driven_qe}}

\section{Related Work}

\subsection{Query Expansion}

Query expansion (QE) is the process of adding relevant terms to a query to improve retrieval performance. In recent years, most QE techniques employed word embeddings~\citep{carpineto2012qesurvey,azad2019qesurvey}.
\citet{kuzi2016query}, \citet{roy2016qe}, \new{and \citet{zamani2016embedding,zamani2016estimating}} used word embeddings to select terms semantically related to the query as expansions.
\citet{diaz2016query} proposed to train embeddings on the top retrieved documents for each query.
\citet{rosin2017learning} leveraged the changing relatedness between words to improve QE for temporal queries.
\citet{zamani2017relevance} proposed relevance-based word embeddings and thus matched the embedding algorithm's objective to that of IR tasks (i.e., to capture relevance).
External data sources (e.g., Wikipedia and WordNet) have also been utilized for QE~\citep{nasir2019knowledge,azad2019qewordnet}.
Recently, \citet{imani2019deep} used a deep neural network classifier to guide the expansion selection process, and~\citet{padaki2020rethinking} showed that using the topic descriptions of TREC collections can improve pretrained reranking models.

In this paper, we perform QE on a specific type of queries, i.e., event-related queries. Moreover, this is the first attempt to leverage events for QE, to the best of our knowledge.

\subsection{Temporal IR and Events}

Temporal information retrieval is a subfield of IR whose purpose is to improve IR by exploiting temporal information in documents and queries~\citep{campos2014surveytir,radinsky2013temporal,kanhabua2016temporal}.
Temporal information has been utilized for query expansion~\citep{rosin2017learning} and query auto-completion~\citep{shokouhi2012time}, among other tasks.
To leverage events for IR tasks, it is often necessary to represent events as continuous vectors (i.e., embeddings).
Most approaches treat events as textual data and apply text embedding methods~\citep{word2vec, wikipedia2vec,rosin2019generating}, while others involve leveraging knowledge graph data~\citep{ding2016knowledge}, or using networks and random walks techniques~\citep{setty2018event2vec}.
The latter has also been used for creating news embeddings~\citep{ma2019news2vec}.

In this work, we embed events using text embedding methods, together with a novel method to combine events and words in a joint vector space.

\subsection{Leveraging Events for Search}

It has been reported that $11\%$ of web search queries are categorized as news~\cite{barilan2009newsqueries}. There have been several attempts to focus on the news domain for search and specifically to leverage events. They all employed supervised learning to detect event-related queries.
\citet{kanhabua2015learning} found query-event candidates using a query log and then determined whether each one is an event using a classifier.
\citet{ghoreishi2013predicting} used a classifier with $20$ features based on top-ranked search results and a query log.
Finally, \citet{zhang2018automatic} published a follow-up study using a similar method with additional features.


All the above studies utilized a query log as the main (if not the only) source of features. In contrast, in our work, we do not rely on query logs but use Wikipedia and DBpedia as sources of events that are open and publicly accessible.

\section{Event and Word Embedding}
\label{sec:embeddings}

To expand an event-related query, we detect the events related to that query, and then, based on the detected events, we derive the candidates for expansion. Both steps require a representation of the words and events.
Previous work has often represented words using a large static corpus, such as Wikipedia~\citep{word2vec,yamada2016joint}. Similarly, events can be represented by the embedding of their corresponding article in Wikipedia~\citep{das2017estimating}. 
We refer to those embeddings as \textit{\static embeddings}.
Alternatively, using \textit{temporal embeddings} allows us to compare words and events over time and focus on the semantics of specific time periods (e.g., given an event, focus on the time period in which it occurred).
We conjecture that utilizing these embeddings would enable us to capture better the interaction between the event and the query expansion candidates.

\subsection{Temporal Embeddings}
\label{sec:temporal_embeddings}

\textit{Temporal word embeddings} were presented in prior work~\citep{hamilton2016diachronic,rosin2017learning,yao2018dynamic,di2019training}. 
The authors created the embeddings using data from a large temporal corpus (e.g., the New York Times (NYT) archive, or Google News). We specifically use the NYT archive.
The embeddings are generated for each time period (i.e., year) and enable us to examine how words meanings and relatedness between words change over time (see Section~\ref{sec:indexing_and_embeddings}).

However, these temporal models may not have meaningful embeddings of events~\citep{rosin2019generating}.
We found only $30\%$ of the event names in our dataset (Section~\ref{sec:data}) to appear (at least once) in their corresponding temporal corpus, and only $12\%$ of them to have at least $10$ occurrences. For example, the string ``Prestige oil spill'' appears once in the NYT corpus of 2002 (when the spill occurred). The search was performed with case ignored, and explicit temporal expressions removed from the event names (e.g., ``1989 Tiananmen Square protests'' was replaced with ``Tiananmen Square protests'').
Moreover, it may take some time until an event's name is determined and referred to in newspapers.
For example, the name ``World War I'' was used only after WWII started.
To conclude, most events do not appear enough times in the temporal corpus to allow their embedding in a vector space. As a result, we are unable to compare events and words.

We therefore present a methodology of representing events in temporal embeddings. We transform each temporal model into a joint vector space for words and events, representing a specific time period. 
We refer to this process as \textit{Event Projection}.

\new{In this work, we use this procedure offline, to enrich the temporal models with events (see implementation details in Section~\ref{sec:indexing_and_embeddings}). We henceforth refer to ``temporal embeddings'' as the enriched temporal embeddings, so that each model of time $t$ would include embeddings of events that occurred in $t$.}


\subsection{Event Projection}
\label{sec:event_projection}
\Static embeddings allow creating a common latent space for \textit{both} words and entities (and specifically, events), allowing us to compare both. 
However, they lack the ability to compare words and events \textit{over time} and focus on the semantics of specific time periods.
We therefore suggest projecting the event \static embeddings into a temporal snapshot.
The event projection method enables us to embed events inside the temporal models. 

\begin{figure}
\centering
\includegraphics[width=0.34\textwidth]{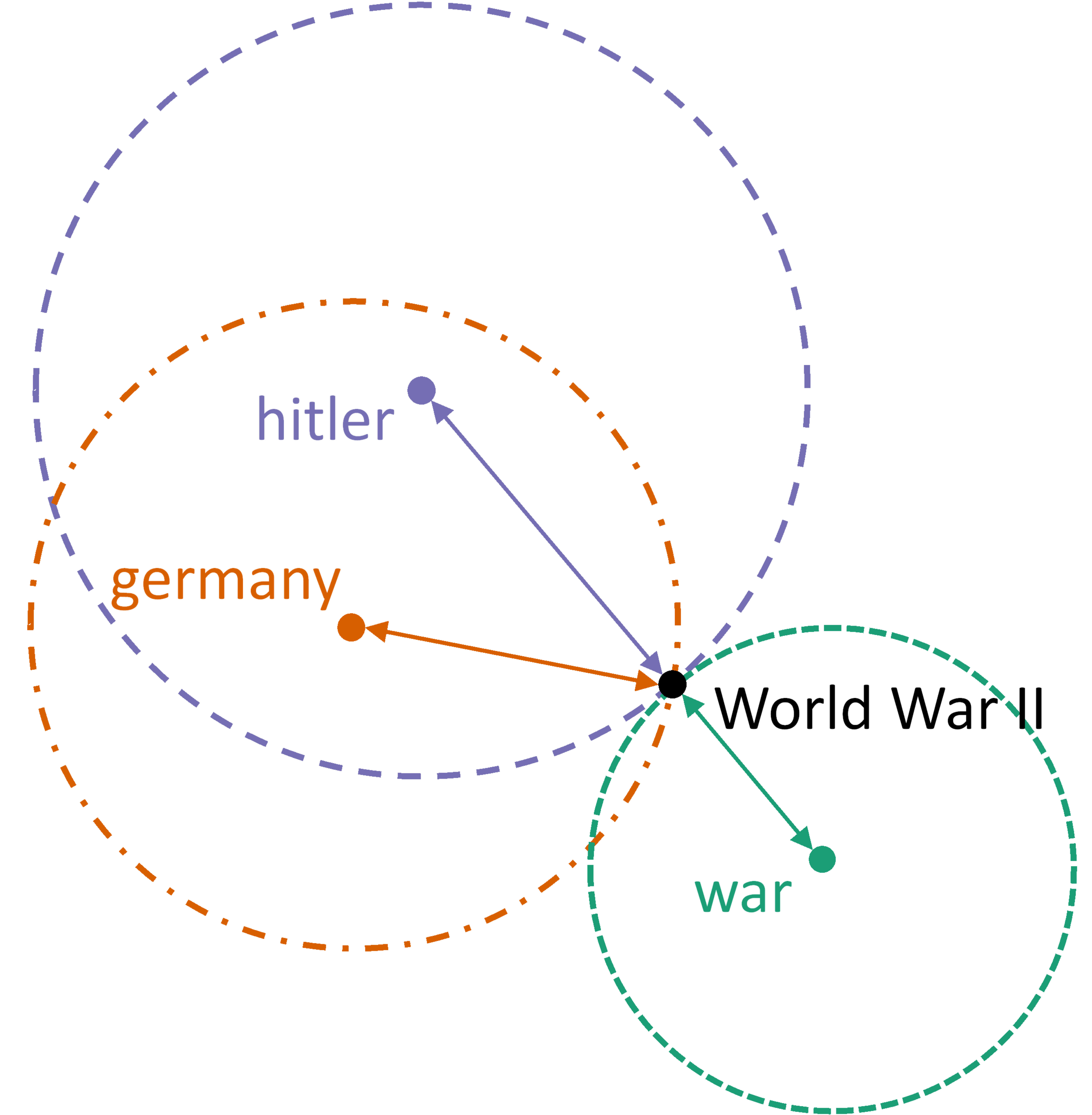}
\caption{\label{fig:triangulation}Trilateration illustration. The position of the event \textit{World War II} can be found by using its distance to the neighboring words \textit{war}, \textit{germany}, and \textit{hitler}.}
\end{figure}

We present a method based on trilateration---a method used to locate an object in space, using measurements to three known objects.
For example, consider the event \textit{World War II}. To project it into a temporal model, we leverage the fact that it is embedded in the \static model.
In that model, we can identify \textit{World War II}'s neighbors (e.g., `germany', `hitler', and `war').
Intuitively, the relationship between \textit{World War II} and its neighbors in the \static model should be similar to that in the temporal model. Exploiting this intuition enables us to position it in the temporal model (Figure~\ref{fig:triangulation}). 

Technically, let $M_t$ be a temporal word embedding model of time $t$, and $e \notin M_t$ be an event. Our goal is to embed $e$ in $M_t$. Algorithm~\ref{alg:event_projection} formally describes the method.

Let $\mathit{Wiki}$ be the \static Wikipedia model. It contains embeddings of events, so $e \in \mathit{Wiki}$.
\new{In line 1,} we take as \textit{anchors} $e$'s top $k$ neighbors in $\mathit{Wiki}$, which also exist in $M_t$.
To identify the neighbors, \new{in line 2} we use a distance function $dist$ to calculate the distance of each anchor $n$ to $e$ in $\mathit{Wiki}$. Specifically, we use cosine distance as $dist$.
Intuitively, since the anchors also exist in $M_t$, we assume that $e$ should be located somewhere between them also in $M_t$, according to the calculated distances.

\new{Technically, finding the best position of $e$ in $M_t$ is an optimization problem, where we seek a position $\hat{v}$ whose distances to the anchors are similar to those of $e$.
Specifically, we minimize the mean squared error (MSE) of the distances of the anchors to $e$ in $\mathit{Wiki}$, and to $v$ in $M_t$ \new{(line 3)}. We use L-BFGS~\citep{byrd1995lbfgs} as the optimization algorithm.}

\begin{algorithm}
	\DontPrintSemicolon
	\KwIn{$M_t$ (source embedding model)}
	\KwIn{$\mathit{Wiki}$ (embedding model of Wikipedia)}
	\KwIn{$e$ (event, $e \notin M_t$)}
	$\mathit{anchors} \gets \{ n : n \in \mathit{kNN_{Wiki}}(e) \ \land \ n \in M_t \}$\;
	$D \gets \{ dist(\mathit{Wiki}(e), \mathit{Wiki}(n)) : n \in \mathit{anchors} \}$\;
	$\hat{v} \gets \argmin_v\{ \mathit{MSE}(D, \{ dist(v, M_t(n)) : n \in \mathit{anchors} \}) \}$\;
	$M_t(e) \gets \hat{v}$\;
	\caption{Event projection algorithm}
	\label{alg:event_projection}
\end{algorithm}

\neww{
Finally, although we described the projection method specifically for our use case for readability, in fact it is a generic method for projecting embeddings from a source model to a target model; it can be applied to any two embedding models, not necessarily Wikipedia and temporal models.
In addition, the projection method does not require the two embedding models to have the same dimensionality, unlike existing mapping methods such as Orthogonal Procrustes~\citep{conneau2017muse}.
}

\section{Event-Related Query Expansion}
\label{sec:method}

Let $q$ be a query defined by a bag of terms $q = \{ w_1, w_2, \dots, w_n \}$.
Our goal is to expand it with \new{semantically related words}. 
Our expansion method is composed of several steps, as depicted in Figure~\ref{fig:pipeline}. We first detect related events to $q$ (Section~\ref{sec:event_detection}), then find expansion candidates based on each event (Section~\ref{sec:event_driven_expansion}), and finally merge them into an expansion for $q$.

We use the following notations throughout the paper:

\begin{definition}
$\cos$ is cosine similarity, which we use as a similarity function between embeddings.
\end{definition}
\begin{definition}
$\cos(w_1, w_2)$ is the similarity between $w_1$ and $w_2$ using the \static model.
\end{definition}
\begin{definition}
$\cos_t(w_1, w_2)$ is the similarity between $w_1$ and $w_2$ during time $t$ (i.e., using the temporal model of time $t$).
\end{definition}
\begin{definition}
$\mathit{tfidf}(w,e)$ is the TF-IDF score of the word $w$ in the event $e$'s Wikipedia entry.
\end{definition}

\subsection{Event Detection}
\label{sec:event_detection}

To detect events related to a query $q$, we start by detecting events for each query term $w{\in}q$ separately, as a set of events, $E_w$. The set is composed of the top-scoring events above a predefined threshold $\mathit{min\_score}$.
\new{We consider two scoring functions for a candidate event $e$}.
The first is embedding-based (called \textit{Similarity}): we calculate the similarity between $w$ and $e$ using a word embedding model.
Technically, it can be either $\cos(w, e)$ if we use the \static model, or $\cos_t(w, e)$ if we use the temporal model of $e$'s time $t$.
The second score is text-based (called \textit{Frequency}): we calculate the frequency of $w$ in $e$'s Wikipedia entry: $\frac{\#w\text{'s occurrences in }e}{\#\text{words in }e}$.
We stem the words and require $w$ itself to appear at least twice in $e$'s entry.

Now, the set of events for $q$ is defined \new{as the set of events that were detected for most query terms:
\begin{align*}
E_q = \{e : e{\in}E_w \text{ for most of } w{\in}q\}
\end{align*}
Finally, to retain only significant events, for every time $t$ we remove from $E_q$ events with a relatively low score. Technically, 
\begin{align*}
E_q = \bigcup_{t} \big\{ e \in E_q^t : \mathit{score}(e) > \mu \cdot \max(\{ \mathit{score}(e) : e \in E_q^t \}) \big\}
\end{align*}
where $\mathit{score}$ is a scoring function, $\mu$ is a parameter, and $E_q^t$ is the set of events that occurred in time $t$}. All parameter values are reported in Section \ref{sec:compared_methods}.

\subsection{Event-Driven Expansion}
\label{sec:event_driven_expansion}

Once related events $E_q$ have been detected, we turn to find expansion terms based on them, as a set $C_q$.
First, for each detected event $e \in E_q$, we create a set of candidate terms, $C_e$.
We examine two types of candidates: terms similar to $e$, and terms similar to $q$.
We use a combination of them to compose $C_e$:
$\lambda k$ top terms by $\mathit{tfidf}(c, e)$, and $(1-\lambda) k$ top terms by $\cos(c, q)$ \new{(where $c$ is a term and $\lambda$ is a parameter)}.

Second, each candidate term $c \in C_e$ is given a score.
We examine two scoring options:

\paragraph{\Static variant} We use the \static embeddings and the following equation to score each candidate:
\begin{align*}
\mathit{score}_S(c) = \alpha \cdot \mathit{tfidf}(c, e) + \beta \cdot \cos(c,e) + \gamma \cdot \cos(e, q)
\end{align*}
\new{where $\alpha$, $\beta$, and $\gamma$ are parameters.}
We refer to our QE method that uses this variant as \textit{\Glob} (for \textit{\textbf{S}tatic \textbf{E}vent-\textbf{D}riven Method}).

\paragraph{Temporal variant} We utilize the temporal embeddings instead of the \static embeddings. This variant works similarly to the \static one, with two differences.
First, the similarities are calculated using temporal embeddings (i.e., given an event $e$ we use the temporal model of $e$'s time $t$).
Second, we add a new feature, called \textit{temporal relevance} ($\mathit{TempRel}$) that prioritizes terms that got closer to $e$'s top neighbors during $e$'s time $t$.
\new{It is calculated as the average amount a candidate $c$ got closer to $e$'s neighborhood:
\begin{align*}
\mathit{TempRel}(c,e) = \frac{1}{k} \sum_{n \in \mathit{kNN}(e)} \frac{\cos_t(c,n)}{\cos_{t-1}(c,n)}
\end{align*}
where $n \in \mathit{kNN}(e)$ and $k$ is a parameter.}
Intuitively, it helps us identify terms whose semantics was influenced by the event.
The equation to score each candidate $c$ is:
\begin{align*}
\mathit{score}_T(c) = & \ \alpha \cdot \mathit{tfidf}(c, e) + \beta \cdot \cos(c,e) \ +  \\
& \ \gamma \cdot \cos(e, q) + \delta \cdot \mathit{TempRel}(c,e)
\end{align*}
\new{where $\alpha$, $\beta$, $\gamma$, and $\lambda$ are parameters.}
We refer to our QE method that uses this variant as \textit{\Tempo} (for \textit{\textbf{T}emporal \textbf{E}vent-\textbf{D}riven Method}).

Finally, the candidates are merged over all events: $C_q = \bigcup_{e \in E_q} C_e$ and the top-scoring terms (by the chosen score from above) are chosen as the actual expansion terms.

\section{Experimental Setup}
\label{sec:evaluation}


All retrieval experiments were carried out using the Terrier search toolkit~\citep{macdonald2012terrier}, with $100$ expansion terms (selected based on a separate validation set, from $\{10, 20, \dots, 150\}$).

\subsection{Event-Related Query Classification}
\label{sec:event_related_query_classification}
We perform our experiments on event-related queries.
To define event relatedness, we consider both explicit relatedness (e.g., ``tropical storms'') and implicit relatedness (e.g., ``police deaths'', due to violent protests).
We present an automated method for identifying such queries. In practice, this will allow search engines in real time to identify queries for which the event-driven QE should be applied.

A query is classified as event-related if most of its terms $w$ have a frequency $\frac{\#w\text{'s occurrences in }e}{\#\text{words in }e} > 0.001$ in any event's Wikipedia entry $e$. 
In addition, we randomly sampled $25\%$ of the queries and manually annotated them by a group of five annotators; they were given various examples of both cases and instructed to mark a query as event-related if they could think of an event that is related to the query, either explicitly or implicitly.
Inter-annotator agreement over all queries, measured using Fleiss's kappa, was $0.76$.
This annotation process deemed that the precision of the method is $68\%$--$76\%$ and the recall is $94\%$--$98\%$ (performance \neww{varies} among the datasets). 

\subsection{Data}
\label{sec:data}

We explore the effectiveness of our proposed QE method on the standard ad-hoc retrieval task using several TREC collections.
\new{Since events are mostly relevant to news, we use newswire datasets. First, we use Robust (TREC Robust Track 2004 collection, which consists of Tipster disks 4 and 5) and TREC12 (consists of Tipster disks 1 and 2).
Second, we use two smaller and more focused datasets, each containing documents from a single newspaper: AP (Associated Press) and WSJ (Wall Street Journal). See Table~\ref{tab:datasets} for details.
Finally, we include in the evaluation an artificial collection of queries (\textit{ALL}), composed of all the queries in the above datasets combined.}
We use the TREC topic titles as queries to the retrieval models. 

\begin{table}
\caption{TREC datasets used for experiments.}
\label{tab:datasets}
\centering
\begin{tabular}{llll}
    \toprule
    Collection & TREC Disks & \# of Docs & \# of Topics\\
    \midrule
    Robust & Disks 4,5-CR & 528,155 & 250\\
    TREC12 & Disks 1--2 & 741,856 & 150\\
    AP & Disks 1--3 & 242,918 & 150\\
    WSJ & Disks 1--2 & 173,252 & 150\\
    \bottomrule
\end{tabular}
\end{table}

\paragraph{Event data} We used DBpedia\footnote{\url{https://wiki.dbpedia.org}} (an open, structured knowledge base derived from Wikipedia) as a source for events and followed the methodology of~\citet{rosin2019generating} to mine events.
We mined \neww{DBpedia entities whose type is `event'}\footnote{i.e., \texttt{yago/Event100029378} or \texttt{Ontology/Event}} and that have an associated Wikipedia entry and year of occurrence.
We retained events corresponding to our focus time period of 1981 to 2018. 
Finally, to reduce the noise of insignificant events, we retained events with over $5000$ monthly views (on average, between 2018--2020) and over $15$ external references.
Our final dataset contains $2354$ events, most related to armed conflicts, politics, disasters, and sports.

\subsection{Indexing and Word Embedding Models}
\label{sec:indexing_and_embeddings}

We indexed the collections using the Terrier search toolkit, with its default setting. We used Porter stemmer for stemming and removed stop-words using the stopping list distributed with Terrier.

For the \static embeddings, we used a pre-trained Wikipedia model from \citet{wikipedia2vec}.
It was created based on the English Wikipedia (dump of April 2018) using word2vec~\citep{mikolov2013distributed}, with a dimensionality of $100$ and a window size of $5$.
It was designed to include embeddings of both words and entities (and specifically, events)~\citep{yamada2016joint}. \new{In particular, the model contains embeddings of all the events in our dataset (Section~\ref{sec:data}).}

For the temporal embeddings, we used the NYT archive\footnote{\url{https://spiderbites.nytimes.com}}, with articles from 1981 to 2018 (11GB of text in total).
For each year of content, we created embeddings using word2vec’s skip-gram with negative sampling, with a window size of $5$ and a dimensionality of $140$, using the Gensim library~\citep{gensim}.
We filtered out words with fewer than $50$ occurrences during that year.
The temporal embeddings were then enriched with events using the projection method described in Section~\ref{sec:temporal_embeddings}.
\new{Each temporal model was enriched with events from our dataset (Section~\ref{sec:data}) that occurred in its time.
For the projection process, we use L-BFGS~\citep{byrd1995lbfgs} as the optimization algorithm, and $k=30$ anchors per event (selected empirically from $\{10,20,30\}$, based on a separate validation set).}

We preferred to use general-purpose corpora (i.e., Wikipedia and NYT) for word embedding, rather than creating embeddings on each specific corpus (as in~\citet{diaz2016query})\neww{, for the sake of generality.}

We created a TF-IDF model of the English Wikipedia based on a dump from March 2020, using the Gensim library.
This model was used by our event-driven QE method (Section~\ref{sec:event_driven_expansion}).

\subsection{Compared Methods}
\label{sec:compared_methods}

\paragraph{Baseline methods}
\new{We consider several baselines to evaluate our methods. First, we compare with three state-of-the-art QE methods (all reimplemented): \citet{imani2019deep}, \citet{roy2016qe}, and \citet{kuzi2016query}.
All these methods leverage word embeddings to yield expansion terms that are semantically close to the query terms. 
Our implementations of these methods use the static Wikipedia embeddings, for a fair comparison.
\citet{imani2019deep} create a deep neural network for classifying expansion terms based on their effectiveness for query expansion, and use it to re-weight expansion terms. \citet{roy2016qe} and \citet{kuzi2016query} look for expansion terms close to the query in a vector space.
\neww{Second, we compare with a BERT~\citep{devlin2018bert} query expansion method, inspired by \citet{padaki2020rethinking}, where the BERT-Base pre-trained model~\citep{devlin2018bert} is used to find expansion terms close to the query.}
Finally, we compare with a standard BM25 approach for IR~\citep{robertson1995okapi}, which does not exploit query expansion techniques.}

We compare the above baselines with our two event-driven methods, i.e., \Glob and \Tempo (Section~\ref{sec:method}).
Both use a standard language model with fixed coefficient interpolation~\citep{zhai2017study}: 
$P(w{\mid}q) = \alpha P_{ED}(w{\mid}q) + (1 - \alpha) P_{ML}(w{\mid}q)$
where $P_{ED}$ is the normalized score of $w$ as calculated by the event-driven method, $P_{ML}$ is the maximum likelihood estimation of $w$, and the interpolation coefficient $\alpha$ is set to $0.6$ (as per the setting prescribed in~\citet{roy2016qe}). We retrieve the documents for the expanded query language model using the TF-IDF weighting model \neww{(we also checked BM25 which showed similar results and trends, so we do not report its results for brevity)}. 

\paragraph{Parameter setting}
\new{The proposed event-driven methods have several hyperparameters.}
To detect events, we set $\mathit{min\_score} = 0.003$, and $\mu = 0.5$ (Section~\ref{sec:event_detection}).
To generate candidate terms per event $e$, we selected $\lambda$ from $\{ 0, 0.1, \dots, 1 \}$.
(Section~\ref{sec:event_driven_expansion}; also see the relevant analysis in Section~\ref{sec:analysis}). 
To score the expansion candidates, we set $\alpha=3, \beta=1, \gamma=1, \delta=1$. For the $\mathit{TempRel}$ feature, we selected $k=5$ nearest neighbors \new{from $\{ 5, 10,20 \}$} (Section~\ref{sec:event_driven_expansion}).
\new{All parameter values are determined based on a separate validation set, where MAP serves as the optimization criterion. To ensure the repeatability of the results, we have released our code and instructions.\footnote{\url{https://github.com/guyrosin/event_driven_qe}}} 

\paragraph{Evaluation metrics} Retrieval effectiveness is measured and compared using several standard metrics~\cite{manning2008introduction}: precision of the top $10$ retrieved documents (P@10), normalized discounted cumulative gain calculated for the top $10$ retrieved documents (NDCG@10), and mean average precision (MAP) of the top-ranked $1000$ documents.
Statistical significance tests were performed using paired t-test at a $95\%$ confidence level.

\section{Results}
\label{sec:results}

\begin{table*}
\caption{Results of query expansion evaluation. Statistically significant differences with BM25, \citet{roy2016qe}, \citet{kuzi2016query}, \citet{imani2019deep}, and BERT baselines are marked with `b', `r', `k', `i', and `B', respectively.}
\label{tab:results}
\centering
\setlength{\tabcolsep}{0.28em}
\begin{tabular}{@{}lccccccccccccccc@{}}
    \toprule
    \multirow{2}[3]{*}{Method} & \multicolumn{3}{c}{Robust} & \multicolumn{3}{c}{TREC12} & \multicolumn{3}{c}{AP} & \multicolumn{3}{c}{WSJ} & \multicolumn{3}{c}{ALL}\\
    \cmidrule(lr){2-4} \cmidrule(lr){5-7} \cmidrule(lr){8-10} \cmidrule(l){11-13} \cmidrule(l){14-16} & 
    P@10 & NDCG & MAP & P@10 & NDCG & MAP & P@10 & NDCG & MAP & P@10 & NDCG & MAP & P@10 & NDCG & MAP\\
	\midrule
	BM25 & 0.40\textss{i} & 0.39 & 0.19 &
	    0.42 & 0.42 & 0.25 &
	    0.29 & 0.30 & 0.20 &
	    0.36 & 0.37 & 0.23 &
	    0.41 & 0.40 & 0.26\\
	Roy \citep{roy2016qe} & 0.38 & 0.38 & 0.19 &
	    0.49\textss{b} & 0.50 & 0.27\textss{b} &
	    0.33 & 0.34 & 0.22\textss{b} &
	    0.37 & 0.38 & 0.25\textss{b} &
	    0.41 & 0.41 & 0.27\\
	Kuzi \citep{kuzi2016query} & 0.36 & 0.37 & 0.19 &
	    0.49 & 0.49 & 0.27\textss{b} &
	    0.35 & 0.33 & 0.21 &
	    0.37 & 0.37 & 0.25\textss{b} &
	    0.39 & 0.40 & 0.27\\
	Imani \citep{imani2019deep} & 0.36 & 0.37 & 0.19 &
	    0.49 & 0.49 & 0.27\textss{b} &
	    0.35 & 0.33 & 0.21 &
	    0.37 & 0.38 & 0.25\textss{b} &
	    0.40 & 0.40 & 0.27\\
	\neww{BERT} & 0.42 & 0.41 & 0.20 &
	    0.46 & 0.46\textss{b} & 0.26\textss{b} &
	    \textbf{0.37}\textss{b} & 0.35 & 0.21 &
	    0.39 & 0.38 & 0.24 &
	    0.42 & 0.43\textss{b} & 0.27\textss{b}\\
	\Tempo & \textbf{0.45}\textss{rki} & \textbf{0.45}\textss{rki} & \textbf{0.23}\textss{brkiB} &
	    \textbf{0.53}\textss{b} & \textbf{0.54}\textss{b} & \textbf{0.31}\textss{brkiB} &
	    \textbf{0.37}\textss{b} & \textbf{0.37}\textss{} & \textbf{0.23}\textss{} &
	    \textbf{0.41}\textss{} & \textbf{0.41}\textss{} & \textbf{0.27}\textss{bB} &
	    \textbf{0.47}\textss{brki} & \textbf{0.48}\textss{brki} & \textbf{0.31}\textss{brkiB}\\
	\bottomrule
\end{tabular}
\end{table*}

In this section, we outline the results of our empirical evaluation.
In all tables throughout the section, the best result in each column is boldfaced, and statistically significant results are marked with `*', unless otherwise specified.
Table~\ref{tab:results} presents the main results of our experiments, comparing our best expansion method \Tempo with the \neww{five} baseline methods.
The baselines perform similarly, while \Tempo outperforms all of them significantly.

\subsection{Analysis}
\label{sec:analysis}

\begin{table*}
\caption{Comparison of embedding types.}
\label{tab:results_embeddings}
\centering
\setlength{\tabcolsep}{0.52em}
\begin{tabular}{@{}lccccccccccccccc@{}}
    \toprule
    \multirow{2}[3]{*}{Method} & \multicolumn{3}{c}{Robust} & \multicolumn{3}{c}{TREC12} & \multicolumn{3}{c}{AP} & \multicolumn{3}{c}{WSJ} & \multicolumn{3}{c}{ALL}\\
    \cmidrule(lr){2-4} \cmidrule(lr){5-7} \cmidrule(lr){8-10} \cmidrule(l){11-13} \cmidrule(l){14-16} & 
    P@10 & NDCG & MAP & P@10 & NDCG & MAP & P@10 & NDCG & MAP & P@10 & NDCG & MAP & P@10 & NDCG & MAP\\
	\midrule
	\Glob & 0.39 & 0.38 & 0.20 & 
	    0.47 & 0.48 & 0.26 &
	    0.33 & 0.36 & 0.22 & 
	    0.36 & 0.37 & 0.25 & 
	    0.42 & 0.41 & 0.27\\
	\Tempo & \textbf{0.45}* & \textbf{0.45} & \textbf{0.23}* & 
	    \textbf{0.53}* & \textbf{0.54} & \textbf{0.31} &
	    \textbf{0.37} & \textbf{0.37} & \textbf{0.23} & 
	    \textbf{0.41} & \textbf{0.41} & \textbf{0.27} & 
	    \textbf{0.47}* & \textbf{0.48} & \textbf{0.31}*\\
	\bottomrule
\end{tabular}
\end{table*}

\neww{We} present a detailed analysis focused on our best performing method \Tempo.

\subsubsection{Embedding type}
We analyze the contribution of various embedding types to event-driven QE. We compare our methods \Glob and \tempo, as the former uses \static embeddings, and the latter uses temporal embeddings.
Table~\ref{tab:results_embeddings} shows the results. The temporal embeddings have a clear advantage in all datasets and metrics.

\subsubsection{Query type}
We compare the performance of \Tempo \neww{with \Glob and the baselines} on different types of queries.
As mentioned in Section~\ref{sec:event_related_query_classification}, event-related queries can be divided into two types: single-event related queries (e.g., ``Schengen agreement'', or ``Tiananmen Square protesters''), and multiple-event related queries (e.g., ``computer viruses'' and ``wrongful convictions'').
We manually created sets of events of either type, from all the TREC datasets used in the evaluation, combined (Section~\ref{sec:data}).
The queries were labeled by a group of three annotators, who were asked to mark the queries as \textit{single-event related}, \textit{multiple-event related} or \textit{N/A}.
The annotators were given guidelines with examples of the different types. Specifically, they were instructed to mark a query as single (multiple)-event related if they can think of a direct connection between the query and a single (multiple) event; and to mark as \textit{N/A} in case the relatedness is vague or non-existing, e.g., for the query ``robotics''. 
$20\%$ and $49\%$ of the event-related queries were labeled as \textit{single-} and \textit{multiple-event related}, respectively; $31\%$ were labeled as \textit{N/A}. 
Inter-annotator agreement, measured using Fleiss' kappa, was $0.78$. 

Table~\ref{tab:comparison_query_type} shows the comparison results. \new{\Tempo is superior for both types of queries. Comparing the event-driven methods, \Tempo outperforms \Glob significantly for multiple-event queries, while for single-event queries it has a much smaller advantage. This is reasonable, as queries that are related to a single event are more specific; their intent is often more focused on the event, so it is relatively easy to identify expansion terms matching the intent.
Comparing the two methods with the baselines strengthens this hypothesis; \Tempo outperforms the baselines consistently for both types of queries, while \Glob is outperformed by the baselines for multiple-event queries.
To conclude, \Tempo shows its advantage on the more difficult type of queries, which is related to multiple events.}

\begin{table}
\caption{Comparison of query types. Statistically significant differences with BM25, \citet{roy2016qe}, \citet{kuzi2016query}, \citet{imani2019deep}, BERT, and \Glob are marked with `b', `r', `k', `i', `B', and `s', respectively.}
\label{tab:comparison_query_type}
\centering
\setlength{\tabcolsep}{0.31em}
\begin{tabular}{@{}lcccccc@{}}
    \toprule
    \multirow{2}[3]{*}{Method} & \multicolumn{3}{c}{Single-Event Queries} & \multicolumn{3}{c}{Multiple-Event Queries}\\
    \cmidrule(lr){2-4} \cmidrule(l){5-7} &
    P@10 & NDCG & MAP & P@10 & NDCG & MAP\\
	\midrule
	\neww{BM25} & 0.46 & 0.37 & 0.38 & 
	0.43 & 0.41 & 0.26\\
	\new{Roy \citep{roy2016qe}} & 0.46 & 0.40\textss{b} & 0.39\textss{b} & 
	0.49\textss{b} & 0.49\textss{b} & 0.29\textss{b}\\
	\new{Kuzi \citep{kuzi2016query}} & 0.45 & 0.39 & 0.39 & 
	0.48 & 0.50\textss{b} & 0.29\textss{b}\\
	\new{Imani \citep{imani2019deep}} & 0.45 & 0.39 & 0.39 & 
	0.50 & 0.50\textss{b} & 0.30\textss{b}\\
	\neww{BERT} & 0.51 & 0.44\textss{b} & 0.41\textss{b} & 
	0.44 & 0.44 & 0.28\\
	\Glob & 0.50 & 0.48\textss{k} & 0.47\textss{bi} & 
	0.46 & 0.46\textss{bB} & 0.25\\
	\Tempo & \textbf{0.52}\textss{ki} & \textbf{0.52}\textss{bk} & \textbf{0.48}\textss{brkiB} & 
	\textbf{0.55}\textss{bsB} & \textbf{0.53}\textss{b} & \textbf{0.31}\textss{brkisB}\\
	\bottomrule
\end{tabular}
\end{table}

\subsubsection{Individual components}
\new{Finally, we analyze the contribution of each of our expansion method's main components: event detection, expansion candidate composition, and expansion candidate scoring.}

\begin{table*}
\caption{Comparison of event detection methods.}
\label{tab:comparison_event_detectors}
\centering
\setlength{\tabcolsep}{0.51em}
\begin{tabular}{@{}lccccccccccccccc@{}}
    \toprule
    \multirow{2}[3]{*}{Method} & \multicolumn{3}{c}{Robust} & \multicolumn{3}{c}{TREC12} & \multicolumn{3}{c}{AP} & \multicolumn{3}{c}{WSJ} & \multicolumn{3}{c}{ALL}\\
    \cmidrule(lr){2-4} \cmidrule(lr){5-7} \cmidrule(lr){8-10} \cmidrule(l){11-13} \cmidrule(l){14-16} & 
    P@10 & NDCG & MAP & P@10 & NDCG & MAP & P@10 & NDCG & MAP & P@10 & NDCG & MAP & P@10 & NDCG & MAP\\
	\midrule
	Sim. & 0.25 & 0.26 & 0.11 & 
	    0.36 & 0.34 & 0.12 & 
	    0.22 & 0.22 & 0.10 & 
	    0.31 & 0.34 & 0.15 & 
	    0.28 & 0.25 & 0.13\\
	Freq. & \textbf{0.45}* & \textbf{0.45}* & \textbf{0.23}* & 
	    \textbf{0.53}* & \textbf{0.54}* & \textbf{0.31}* &
	    \textbf{0.37} & \textbf{0.37} & \textbf{0.23}* & 
	    \textbf{0.41} & \textbf{0.41} & \textbf{0.27}* & 
	    \textbf{0.47}* & \textbf{0.43}* & \textbf{0.31}*\\
	\bottomrule
\end{tabular}
\end{table*}

\paragraph{Event detection method}
In Section~\ref{sec:event_detection}, we suggest two methods for related event detection, namely \textit{Similarity} and \textit{Frequency}.
For comparing them, we execute \Tempo with each. Table~\ref{tab:comparison_event_detectors} shows the results: the \textit{Frequency} method performs better for all datasets (significantly, in most cases).
We believe the reason for this difference is that word2vec similarity is not precise enough for this task: we observed many false positive events using the \textit{Similarity} method. In contrast, the \textit{Frequency} method guarantees that the detected events include the query words (frequently) in their Wikipedia entry, and is therefore more precise.

\begin{figure}
\centering
\includegraphics[width=0.7\linewidth]{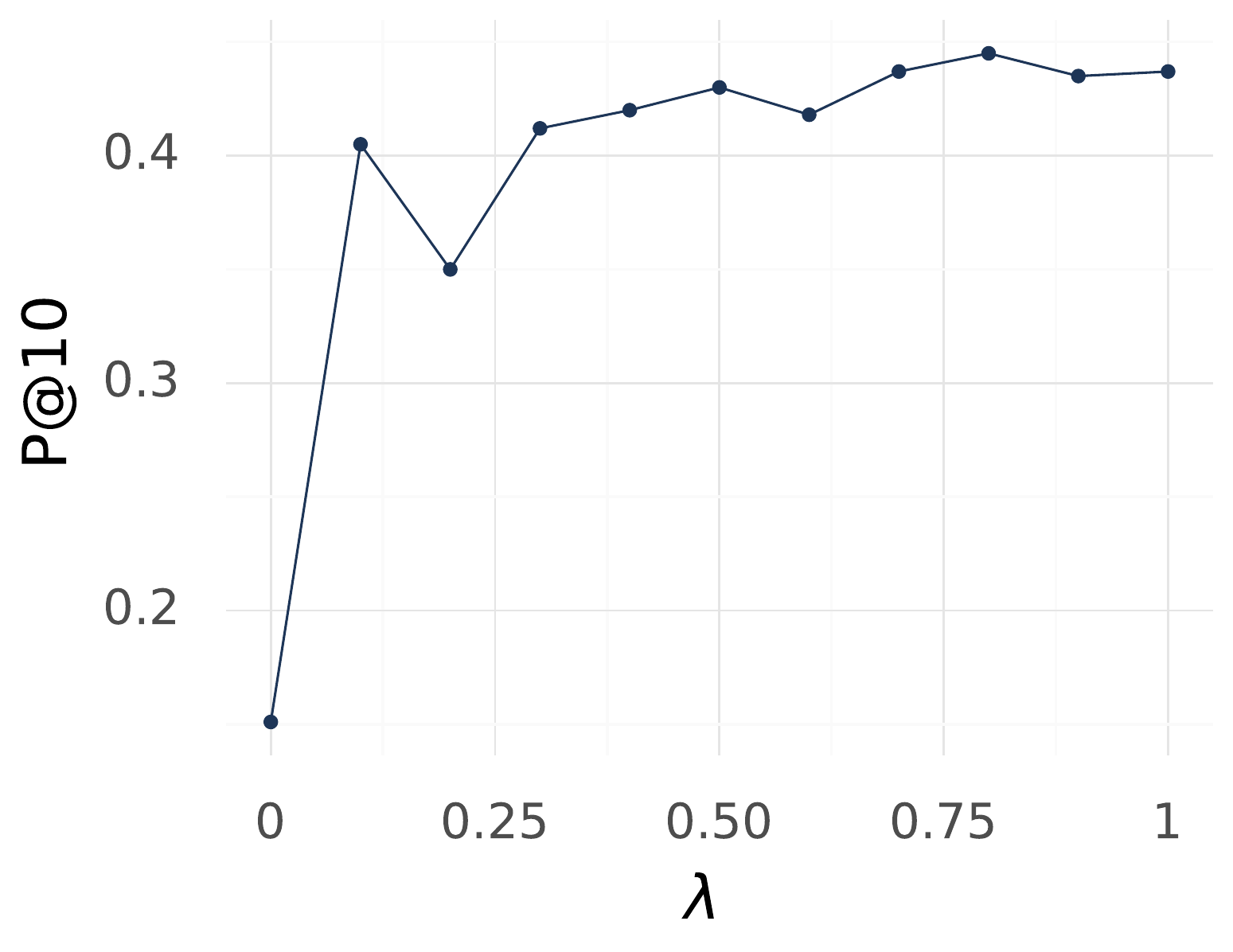}
\caption{Sensitivity of TED to the interpolation parameter $\lambda$ (Section~\ref{sec:event_driven_expansion}) for the Robust dataset.}
\label{fig:candidates_lambda_tuning}
\end{figure}

\paragraph{Expansion candidate composition}
We compare different compositions of expansion candidates, given a query $q$ and a related event $e$ (Section~\ref{sec:event_driven_expansion}).
Figure~\ref{fig:candidates_lambda_tuning} plots the sensitivity of \Tempo to the interpolation parameter $\lambda$ for the Robust dataset (other datasets behave similarly). $\lambda$ controls the balance between candidates similar to $q$ ($\lambda=0$) and candidates related to $e$ ($\lambda=1$).
The plot has a positive trend, and the optimal $\lambda$ is $0.8$.
We deduce that: (1) terms related to the event are more beneficial to the query expansion than terms similar to the query; (2) using a small number of terms similar to the query is helpful;
it diversifies the expansion and acts as a fallback in case of a false positive (i.e., an unrelated event was detected).

\begin{table*}
\centering
\caption{Comparison of expansion candidate scoring features in a leave-one-out setting.}
\label{tab:comparison_candidate_scoring}
\begin{tabular}{@{}lcccccccccccc@{}}
    \toprule
    \multirow{2}[3]{*}{Left-Out Feature} & \multicolumn{3}{c}{Robust} & \multicolumn{3}{c}{TREC12} & \multicolumn{3}{c}{AP} & \multicolumn{3}{c}{WSJ}\\
    \cmidrule(lr){2-4} \cmidrule(lr){5-7} \cmidrule(lr){8-10} \cmidrule(l){11-13} & 
    P@10 & NDCG & MAP & P@10 & NDCG & MAP & P@10 & NDCG & MAP & P@10 & NDCG & MAP\\
	\midrule
	$\cos(c,q)$ & 0.44 & 0.45 & \textbf{0.24} &
	    0.49 & 0.51 & 0.29 & 
	    0.35 & 0.35 & 0.22 & 
	    0.35 & 0.36 & 0.25 \\
	$\cos(c,e)$ & \textbf{0.46} & \textbf{0.46} & 0.23 &
	    \textbf{0.53} & \textbf{0.54} & \textbf{0.31} & 
	    \textbf{0.38} & \textbf{0.38} & \textbf{0.23} & 
	    \textbf{0.41} & 0.41 & \textbf{0.27} \\
	$\cos(e,q)$ & 0.45 & 0.45 & 0.23 &
	    0.51 & 0.53 & \textbf{0.31} & 
	    0.35 & 0.36 & \textbf{0.23} & 
	    \textbf{0.41} & 0.41 & \textbf{0.27} \\
	$\mathit{tfidf}(c,e)$ & 0.45 & 0.45 & \textbf{0.23} &
	    \textbf{0.53} & \textbf{0.54} & \textbf{0.31} & 
	    0.37 & 0.37 & 0.23 & 
	    \textbf{0.41} & \textbf{0.42} & \textbf{0.27} \\
	$\mathit{TempRel}(c,e)$ & 0.45 & 0.45 & \textbf{0.23} &
	    0.52 & 0.53 & \textbf{0.31} & 
	    0.35 & 0.36 & \textbf{0.23} & 
	    \textbf{0.41} & 0.41 & \textbf{0.27} \\
	\midrule
	All features & 0.45 & 0.45 & 0.23 &
	    0.53 & \textbf{0.54} & \textbf{0.31} & 
	    0.37 & 0.37 & \textbf{0.23} & 
	    \textbf{0.41} & 0.41 & \textbf{0.27}\\
	\bottomrule
\end{tabular}
\end{table*}

\paragraph{Expansion candidate scoring}
Table~\ref{tab:comparison_candidate_scoring} presents a leave-one-out comparison of the features used to score expansion candidates by \Tempo: term-query similarity, term-event similarity, event-query similarity, term-event TF-IDF score, and temporal relevance (Section~\ref{sec:event_driven_expansion}).
We observe an almost equal performance, i.e., no single feature \neww{is more important than any other}. \new{Term-query similarity is relatively strong, and on the other hand term-event similarity is relatively weak, but the differences are not statistically significant.}

To conclude, the contributing factors to \Tempo's performance are detecting the right events and using the right embeddings. In contrast, the choice of candidate scoring features is insignificant.

\subsection{Qualitative Examples}
\label{sec:examples}
\new{To gain more intuition on when and why event-driven QE works, we provide examples with queries from the Robust and TREC12 datasets.
For each query, we show the top detected events by \Tempo, top five expansions produced by it, and for comparison---top five expansions produced by a baseline~\citep{imani2019deep}.}

\paragraph{\new{Positive examples}}
\new{Table~\ref{tab:positive_examples} shows successful examples of expansions made by \Tempo.
For the query ``African civilian deaths'', the baseline produced as expansions names of African nationalities, while \Tempo successfully detected several civil wars that occurred in Africa and produced terms related to them.}

\new{Similarly, for the query ``U.S. invasion of Panama'', the baseline produced mainly countries or terms similar to Panama. In contrast, \Tempo produced expansion terms that are more related to the invasion itself, which is closer to the query intent.}

\new{The third example, ``Tiananmen Square protesters'', is easier to expand. The expansions of both methods are similarly focused on the word `protesters'.}

\paragraph{\new{Negative examples}}
\new{Table~\ref{tab:negative_examples} demonstrates errors made by \Tempo.
The first type of error is detecting wrong events. For example, given the query ``Bank Failures'', the detected event ``2000 Camp David Summit'' is not relevant to the query, but was detected nonetheless because (1) the Summit is considered a failure, and (2) the phrase ``West Bank'' appears frequently in the event's Wikipedia entry. It is interesting to note that despite the wrong detection, the expansions produced by \Tempo are mostly reasonable; just one term out of the top five expansions is not related to the query (`Israel').}

\new{Another error \Tempo can make is to produce wrong expansions, based on a relevant event. For the query ``Death from Cancer'', the baseline produced reasonable cancer-related expansions. \Tempo detects Chernobyl Disaster, which is a relevant event. The expansions are reasonable, but since only one event was detected, the focus on it is considerable; two out of the top five expansions are more relevant to the event than to the query (`reactor' and `dose').}

\new{Similarly, for the query ``Greenpeace'', the baseline produced reasonable expansions related to the environmental organization. \Tempo detected a relevant event (a bombing operation of Greenpeace's flagship), but the expansions are too focused on the event, and so they are less relevant to the query (e.g., `bomb', `yacht', and `warrior').
}

\begin{table*}
\caption{\new{Positive examples of expansions created by \Tempo, compared with a baseline \citep{imani2019deep}.}}
\label{tab:positive_examples}
\centering
\begin{tabular}{@{}L{37mm}L{40mm}L{35mm}L{42mm}@{}}
    \toprule
    Query & Top Detected Events & Top Expansions & Baseline's Top Expansions\\
    \midrule
    African civilian deaths & War in Darfur, Somali Civil War, Sudanese Civil War & uganda, sudan, military, force, tutsi & sudanese, rhodesian, ugandan, rwandan, namibian\\[1.1em]
    U.S. invasion of Panama & US invasion of Panama & invade, air, operation, occupation, force & nicaragua, reoccupation, guatemala, ryukyus, hispaniola \\[1.1em]
    Tiananmen Square protesters & 1989 Tiananmen Square protests & demonstration, crowd, riot, beijing, protests & protesters, demonstrators, protesting, marchers, protests \\
    \bottomrule
\end{tabular}
\end{table*}

\begin{table*}
\caption{\new{Negative examples of expansions created by \Tempo, compared with a baseline \citep{imani2019deep}.}}
\label{tab:negative_examples}
\centering
\begin{tabular}{@{}L{37mm}L{40mm}L{35mm}L{42mm}@{}}
    \toprule
    Query & Top Detected Events & Top Expansions & Baseline's Top Expansions\\
    \midrule
    \new{Bank Failures} & 2000 Camp David Summit & loan, deposit, israel, banks, lending & banks, collapse, depositors, failures, savings \\[1.1em]
    \new{Death from Cancer} & Chernobyl Disaster & disease, illness, lymph, reactor, dose & leukemia, prostate, pancreatic, melanoma, tumour \\[1.1em]
    \new{Greenpeace} & Sinking of the Rainbow Warrior & bomb, environmentalist, yacht, auckland, warrior & environmentalists, avaaz worldwatch, actionaid, adbusters \\
    \bottomrule
\end{tabular}
\end{table*}

\section{Conclusions}
\label{sec:conclusions}

In this paper, we presented a novel approach of event-driven query expansion. 
Our approach identifies events related to the query and then suggests expansion terms that are semantically related to those events or influenced by them.
We studied different embedding types for words and events. We identified that temporal embeddings coupled with a mechanism to simultaneously embed words and events in the same space significantly improve QE performance. The mechanism operates by projecting event embeddings from an auxiliary model (Wikipedia) to temporal models. This enables us to compare words and events at specific times.
We analyzed different event-related queries and conclude that temporal embeddings significantly improve multiple-event queries. Additionally, we studied several methods for event detection, expansion candidate composition, and expansion candidate scoring.
We showed that our proposed method improves QE performance significantly on various \new{newswire} TREC collections, compared with state-of-the-art baselines.
For future work, we intend to leverage the event-driven methodology for other IR tasks, such as reranking and query classification. 

\bibliographystyle{ACM-Reference-Format}
\balance
\bibliography{main}


\begin{thebibliography}{46}


\ifx \showCODEN    \undefined \def \showCODEN     #1{\unskip}     \fi
\ifx \showDOI      \undefined \def \showDOI       #1{#1}\fi
\ifx \showISBNx    \undefined \def \showISBNx     #1{\unskip}     \fi
\ifx \showISBNxiii \undefined \def \showISBNxiii  #1{\unskip}     \fi
\ifx \showISSN     \undefined \def \showISSN      #1{\unskip}     \fi
\ifx \showLCCN     \undefined \def \showLCCN      #1{\unskip}     \fi
\ifx \shownote     \undefined \def \shownote      #1{#1}          \fi
\ifx \showarticletitle \undefined \def \showarticletitle #1{#1}   \fi
\ifx \showURL      \undefined \def \showURL       {\relax}        \fi
\providecommand\bibfield[2]{#2}
\providecommand\bibinfo[2]{#2}
\providecommand\natexlab[1]{#1}
\providecommand\showeprint[2][]{arXiv:#2}

\bibitem[\protect\citeauthoryear{Amati and Van~Rijsbergen}{Amati and
  Van~Rijsbergen}{2002}]%
        {amati2002probabilistic}
\bibfield{author}{\bibinfo{person}{Gianni Amati} {and}
  \bibinfo{person}{Cornelis~Joost Van~Rijsbergen}.}
  \bibinfo{year}{2002}\natexlab{}.
\newblock \showarticletitle{Probabilistic models of information retrieval based
  on measuring the divergence from randomness}.
\newblock \bibinfo{journal}{\emph{ACM Transactions on Information Systems
  (TOIS)}} \bibinfo{volume}{20}, \bibinfo{number}{4} (\bibinfo{year}{2002}),
  \bibinfo{pages}{357--389}.
\newblock


\bibitem[\protect\citeauthoryear{Azad and Deepak}{Azad and Deepak}{2019a}]%
        {azad2019qewordnet}
\bibfield{author}{\bibinfo{person}{Hiteshwar~Kumar Azad} {and}
  \bibinfo{person}{Akshay Deepak}.} \bibinfo{year}{2019}\natexlab{a}.
\newblock \showarticletitle{A new approach for query expansion using Wikipedia
  and WordNet}.
\newblock \bibinfo{journal}{\emph{Information Sciences}}  \bibinfo{volume}{492}
  (\bibinfo{year}{2019}), \bibinfo{pages}{147--163}.
\newblock


\bibitem[\protect\citeauthoryear{Azad and Deepak}{Azad and Deepak}{2019b}]%
        {azad2019qesurvey}
\bibfield{author}{\bibinfo{person}{Hiteshwar~Kumar Azad} {and}
  \bibinfo{person}{Akshay Deepak}.} \bibinfo{year}{2019}\natexlab{b}.
\newblock \showarticletitle{Query expansion techniques for information
  retrieval: A survey}.
\newblock \bibinfo{journal}{\emph{Information Processing \& Management}}
  \bibinfo{volume}{56}, \bibinfo{number}{5} (\bibinfo{year}{2019}),
  \bibinfo{pages}{1698--1735}.
\newblock


\bibitem[\protect\citeauthoryear{Bar-Ilan, Zhu, and Levene}{Bar-Ilan
  et~al\mbox{.}}{2009}]%
        {barilan2009newsqueries}
\bibfield{author}{\bibinfo{person}{Judit Bar-Ilan}, \bibinfo{person}{Zheng
  Zhu}, {and} \bibinfo{person}{Mark Levene}.} \bibinfo{year}{2009}\natexlab{}.
\newblock \showarticletitle{Topic-Specific Analysis of Search Queries}. In
  \bibinfo{booktitle}{\emph{Proceedings of the 2009 Workshop on Web Search
  Click Data}} (Barcelona, Spain) \emph{(\bibinfo{series}{WSCD ’09})}.
  \bibinfo{publisher}{Association for Computing Machinery},
  \bibinfo{address}{New York, NY, USA}, \bibinfo{pages}{35–42}.
\newblock
\showISBNx{9781605584348}
\urldef\tempurl%
\url{https://doi.org/10.1145/1507509.1507515}
\showDOI{\tempurl}


\bibitem[\protect\citeauthoryear{Byrd, Lu, Nocedal, and Zhu}{Byrd
  et~al\mbox{.}}{1995}]%
        {byrd1995lbfgs}
\bibfield{author}{\bibinfo{person}{Richard~H Byrd}, \bibinfo{person}{Peihuang
  Lu}, \bibinfo{person}{Jorge Nocedal}, {and} \bibinfo{person}{Ciyou Zhu}.}
  \bibinfo{year}{1995}\natexlab{}.
\newblock \showarticletitle{A limited memory algorithm for bound constrained
  optimization}.
\newblock \bibinfo{journal}{\emph{SIAM Journal on scientific computing}}
  \bibinfo{volume}{16}, \bibinfo{number}{5} (\bibinfo{year}{1995}),
  \bibinfo{pages}{1190--1208}.
\newblock


\bibitem[\protect\citeauthoryear{Campos, Dias, Jorge, and Jatowt}{Campos
  et~al\mbox{.}}{2014}]%
        {campos2014surveytir}
\bibfield{author}{\bibinfo{person}{Ricardo Campos}, \bibinfo{person}{Ga{\"e}l
  Dias}, \bibinfo{person}{Al{\'\i}pio~M Jorge}, {and} \bibinfo{person}{Adam
  Jatowt}.} \bibinfo{year}{2014}\natexlab{}.
\newblock \showarticletitle{Survey of temporal information retrieval and
  related applications}.
\newblock \bibinfo{journal}{\emph{ACM Computing Surveys (CSUR)}}
  \bibinfo{volume}{47}, \bibinfo{number}{2} (\bibinfo{year}{2014}),
  \bibinfo{pages}{1--41}.
\newblock


\bibitem[\protect\citeauthoryear{Carpineto and Romano}{Carpineto and
  Romano}{2012}]%
        {carpineto2012qesurvey}
\bibfield{author}{\bibinfo{person}{Claudio Carpineto} {and}
  \bibinfo{person}{Giovanni Romano}.} \bibinfo{year}{2012}\natexlab{}.
\newblock \showarticletitle{A Survey of Automatic Query Expansion in
  Information Retrieval}.
\newblock \bibinfo{journal}{\emph{ACM Comput. Surv.}} \bibinfo{volume}{44},
  \bibinfo{number}{1} (\bibinfo{date}{Jan.} \bibinfo{year}{2012}).
\newblock


\bibitem[\protect\citeauthoryear{Conneau, Lample, Ranzato, Denoyer, and
  J{\'e}gou}{Conneau et~al\mbox{.}}{2017}]%
        {conneau2017muse}
\bibfield{author}{\bibinfo{person}{Alexis Conneau}, \bibinfo{person}{Guillaume
  Lample}, \bibinfo{person}{Marc'Aurelio Ranzato}, \bibinfo{person}{Ludovic
  Denoyer}, {and} \bibinfo{person}{Herv{\'e} J{\'e}gou}.}
  \bibinfo{year}{2017}\natexlab{}.
\newblock \showarticletitle{Word Translation Without Parallel Data}.
\newblock \bibinfo{journal}{\emph{arXiv preprint arXiv:1710.04087}}
  (\bibinfo{year}{2017}).
\newblock


\bibitem[\protect\citeauthoryear{Das, Mishra, Berberich, and Setty}{Das
  et~al\mbox{.}}{2017}]%
        {das2017estimating}
\bibfield{author}{\bibinfo{person}{Supratim Das}, \bibinfo{person}{Arunav
  Mishra}, \bibinfo{person}{Klaus Berberich}, {and} \bibinfo{person}{Vinay
  Setty}.} \bibinfo{year}{2017}\natexlab{}.
\newblock \showarticletitle{Estimating event focus time using neural word
  embeddings}. In \bibinfo{booktitle}{\emph{Proceedings of the 2017 ACM on
  Conference on Information and Knowledge Management}}.
  \bibinfo{pages}{2039--2042}.
\newblock


\bibitem[\protect\citeauthoryear{Devlin, Chang, Lee, and Toutanova}{Devlin
  et~al\mbox{.}}{2018}]%
        {devlin2018bert}
\bibfield{author}{\bibinfo{person}{Jacob Devlin}, \bibinfo{person}{Ming-Wei
  Chang}, \bibinfo{person}{Kenton Lee}, {and} \bibinfo{person}{Kristina
  Toutanova}.} \bibinfo{year}{2018}\natexlab{}.
\newblock \showarticletitle{Bert: Pre-training of deep bidirectional
  transformers for language understanding}.
\newblock \bibinfo{journal}{\emph{arXiv preprint arXiv:1810.04805}}
  (\bibinfo{year}{2018}).
\newblock


\bibitem[\protect\citeauthoryear{Di~Carlo, Bianchi, and Palmonari}{Di~Carlo
  et~al\mbox{.}}{2019}]%
        {di2019training}
\bibfield{author}{\bibinfo{person}{Valerio Di~Carlo}, \bibinfo{person}{Federico
  Bianchi}, {and} \bibinfo{person}{Matteo Palmonari}.}
  \bibinfo{year}{2019}\natexlab{}.
\newblock \showarticletitle{Training temporal word embeddings with a compass}.
  In \bibinfo{booktitle}{\emph{Proceedings of the AAAI Conference on Artificial
  Intelligence}}, Vol.~\bibinfo{volume}{33}. \bibinfo{pages}{6326--6334}.
\newblock


\bibitem[\protect\citeauthoryear{Diaz, Mitra, and Craswell}{Diaz
  et~al\mbox{.}}{2016}]%
        {diaz2016query}
\bibfield{author}{\bibinfo{person}{Fernando Diaz}, \bibinfo{person}{Bhaskar
  Mitra}, {and} \bibinfo{person}{Nick Craswell}.}
  \bibinfo{year}{2016}\natexlab{}.
\newblock \showarticletitle{Query expansion with locally-trained word
  embeddings}.
\newblock \bibinfo{journal}{\emph{arXiv preprint arXiv:1605.07891}}
  (\bibinfo{year}{2016}).
\newblock


\bibitem[\protect\citeauthoryear{Ding, Zhang, Liu, and Duan}{Ding
  et~al\mbox{.}}{2016}]%
        {ding2016knowledge}
\bibfield{author}{\bibinfo{person}{Xiao Ding}, \bibinfo{person}{Yue Zhang},
  \bibinfo{person}{Ting Liu}, {and} \bibinfo{person}{Junwen Duan}.}
  \bibinfo{year}{2016}\natexlab{}.
\newblock \showarticletitle{Knowledge-driven event embedding for stock
  prediction}. In \bibinfo{booktitle}{\emph{Proceedings of coling 2016, the
  26th international conference on computational linguistics: Technical
  papers}}. \bibinfo{pages}{2133--2142}.
\newblock


\bibitem[\protect\citeauthoryear{Gabrilovich, Markovitch,
  et~al\mbox{.}}{Gabrilovich et~al\mbox{.}}{2007}]%
        {gabrilovich2007computing}
\bibfield{author}{\bibinfo{person}{Evgeniy Gabrilovich}, \bibinfo{person}{Shaul
  Markovitch}, {et~al\mbox{.}}} \bibinfo{year}{2007}\natexlab{}.
\newblock \showarticletitle{Computing semantic relatedness using
  wikipedia-based explicit semantic analysis.}. In
  \bibinfo{booktitle}{\emph{IJcAI}}, Vol.~\bibinfo{volume}{7}.
  \bibinfo{pages}{1606--1611}.
\newblock


\bibitem[\protect\citeauthoryear{Ghoreishi and Sun}{Ghoreishi and Sun}{2013}]%
        {ghoreishi2013predicting}
\bibfield{author}{\bibinfo{person}{Seyyedeh~Newsha Ghoreishi} {and}
  \bibinfo{person}{Aixin Sun}.} \bibinfo{year}{2013}\natexlab{}.
\newblock \showarticletitle{Predicting event-relatedness of popular queries}.
  In \bibinfo{booktitle}{\emph{Proceedings of the 22nd ACM international
  conference on Information \& Knowledge Management}}.
  \bibinfo{pages}{1193--1196}.
\newblock


\bibitem[\protect\citeauthoryear{Hamilton, Leskovec, and Jurafsky}{Hamilton
  et~al\mbox{.}}{2016}]%
        {hamilton2016diachronic}
\bibfield{author}{\bibinfo{person}{William~L Hamilton}, \bibinfo{person}{Jure
  Leskovec}, {and} \bibinfo{person}{Dan Jurafsky}.}
  \bibinfo{year}{2016}\natexlab{}.
\newblock \showarticletitle{Diachronic Word Embeddings Reveal Statistical Laws
  of Semantic Change}. In \bibinfo{booktitle}{\emph{Proceedings of the 54th
  Annual Meeting of the Association for Computational Linguistics (Volume 1:
  Long Papers)}}, Vol.~\bibinfo{volume}{1}. \bibinfo{pages}{1489--1501}.
\newblock


\bibitem[\protect\citeauthoryear{Imani, Vakili, Montazer, and Shakery}{Imani
  et~al\mbox{.}}{2019}]%
        {imani2019deep}
\bibfield{author}{\bibinfo{person}{Ayyoob Imani}, \bibinfo{person}{Amir
  Vakili}, \bibinfo{person}{Ali Montazer}, {and} \bibinfo{person}{Azadeh
  Shakery}.} \bibinfo{year}{2019}\natexlab{}.
\newblock \showarticletitle{Deep neural networks for query expansion using word
  embeddings}. In \bibinfo{booktitle}{\emph{European Conference on Information
  Retrieval}}. Springer, \bibinfo{pages}{203--210}.
\newblock


\bibitem[\protect\citeauthoryear{Kanhabua and Anand}{Kanhabua and
  Anand}{2016}]%
        {kanhabua2016temporal}
\bibfield{author}{\bibinfo{person}{Nattiya Kanhabua} {and}
  \bibinfo{person}{Avishek Anand}.} \bibinfo{year}{2016}\natexlab{}.
\newblock \showarticletitle{Temporal information retrieval}. In
  \bibinfo{booktitle}{\emph{Proceedings of the 39th International ACM SIGIR
  conference on Research and Development in Information Retrieval}}.
  \bibinfo{pages}{1235--1238}.
\newblock


\bibitem[\protect\citeauthoryear{Kanhabua, Ngoc~Nguyen, and Nejdl}{Kanhabua
  et~al\mbox{.}}{2015}]%
        {kanhabua2015learning}
\bibfield{author}{\bibinfo{person}{Nattiya Kanhabua}, \bibinfo{person}{Tu
  Ngoc~Nguyen}, {and} \bibinfo{person}{Wolfgang Nejdl}.}
  \bibinfo{year}{2015}\natexlab{}.
\newblock \showarticletitle{Learning to detect event-related queries for web
  search}. In \bibinfo{booktitle}{\emph{Proceedings of the 24th International
  Conference on World Wide Web}}. \bibinfo{pages}{1339--1344}.
\newblock


\bibitem[\protect\citeauthoryear{Kanhabua and N{\o}rv{\aa}g}{Kanhabua and
  N{\o}rv{\aa}g}{2010}]%
        {kanhabua2010determining}
\bibfield{author}{\bibinfo{person}{Nattiya Kanhabua} {and}
  \bibinfo{person}{Kjetil N{\o}rv{\aa}g}.} \bibinfo{year}{2010}\natexlab{}.
\newblock \showarticletitle{Determining time of queries for re-ranking search
  results}. In \bibinfo{booktitle}{\emph{International Conference on Theory and
  Practice of Digital Libraries}}. Springer, \bibinfo{pages}{261--272}.
\newblock


\bibitem[\protect\citeauthoryear{Kuzi, Shtok, and Kurland}{Kuzi
  et~al\mbox{.}}{2016}]%
        {kuzi2016query}
\bibfield{author}{\bibinfo{person}{Saar Kuzi}, \bibinfo{person}{Anna Shtok},
  {and} \bibinfo{person}{Oren Kurland}.} \bibinfo{year}{2016}\natexlab{}.
\newblock \showarticletitle{Query expansion using word embeddings}. In
  \bibinfo{booktitle}{\emph{Proceedings of the 25th ACM international on
  conference on information and knowledge management}}.
  \bibinfo{pages}{1929--1932}.
\newblock


\bibitem[\protect\citeauthoryear{Ma, Zong, Yang, and Su}{Ma
  et~al\mbox{.}}{2019}]%
        {ma2019news2vec}
\bibfield{author}{\bibinfo{person}{Ye Ma}, \bibinfo{person}{Lu Zong},
  \bibinfo{person}{Yikang Yang}, {and} \bibinfo{person}{Jionglong Su}.}
  \bibinfo{year}{2019}\natexlab{}.
\newblock \showarticletitle{News2vec: News Network Embedding with Subnode
  Information}. In \bibinfo{booktitle}{\emph{Proceedings of the 2019 Conference
  on Empirical Methods in Natural Language Processing and the 9th International
  Joint Conference on Natural Language Processing (EMNLP-IJCNLP)}}.
  \bibinfo{pages}{4845--4854}.
\newblock


\bibitem[\protect\citeauthoryear{Macdonald, McCreadie, Santos, and
  Ounis}{Macdonald et~al\mbox{.}}{2012}]%
        {macdonald2012terrier}
\bibfield{author}{\bibinfo{person}{Craig Macdonald}, \bibinfo{person}{Richard
  McCreadie}, \bibinfo{person}{Rodrygo~LT Santos}, {and} \bibinfo{person}{Iadh
  Ounis}.} \bibinfo{year}{2012}\natexlab{}.
\newblock \showarticletitle{From puppy to maturity: Experiences in developing
  Terrier}.
\newblock \bibinfo{journal}{\emph{Proc. of OSIR at SIGIR}}
  (\bibinfo{year}{2012}), \bibinfo{pages}{60--63}.
\newblock


\bibitem[\protect\citeauthoryear{Manning, Raghavan, and Sch{\"u}tze}{Manning
  et~al\mbox{.}}{2008}]%
        {manning2008introduction}
\bibfield{author}{\bibinfo{person}{Christopher~D Manning},
  \bibinfo{person}{Prabhakar Raghavan}, {and} \bibinfo{person}{Hinrich
  Sch{\"u}tze}.} \bibinfo{year}{2008}\natexlab{}.
\newblock \bibinfo{booktitle}{\emph{Introduction to information retrieval}}.
\newblock \bibinfo{publisher}{Cambridge university press}.
\newblock


\bibitem[\protect\citeauthoryear{Mikolov, Sutskever, Chen, Corrado, and
  Dean}{Mikolov et~al\mbox{.}}{2013a}]%
        {word2vec}
\bibfield{author}{\bibinfo{person}{Tomas Mikolov}, \bibinfo{person}{Ilya
  Sutskever}, \bibinfo{person}{Kai Chen}, \bibinfo{person}{Greg~S Corrado},
  {and} \bibinfo{person}{Jeff Dean}.} \bibinfo{year}{2013}\natexlab{a}.
\newblock \showarticletitle{Distributed representations of words and phrases
  and their compositionality}. In \bibinfo{booktitle}{\emph{Advances in neural
  information processing systems}}. \bibinfo{pages}{3111--3119}.
\newblock


\bibitem[\protect\citeauthoryear{Mikolov, Sutskever, Chen, Corrado, and
  Dean}{Mikolov et~al\mbox{.}}{2013b}]%
        {mikolov2013distributed}
\bibfield{author}{\bibinfo{person}{Tomas Mikolov}, \bibinfo{person}{Ilya
  Sutskever}, \bibinfo{person}{Kai Chen}, \bibinfo{person}{Greg~S Corrado},
  {and} \bibinfo{person}{Jeff Dean}.} \bibinfo{year}{2013}\natexlab{b}.
\newblock \showarticletitle{Distributed representations of words and phrases
  and their compositionality}. In \bibinfo{booktitle}{\emph{Advances in neural
  information processing systems}}. \bibinfo{pages}{3111--3119}.
\newblock


\bibitem[\protect\citeauthoryear{Nasir, Varlamis, and Ishfaq}{Nasir
  et~al\mbox{.}}{2019}]%
        {nasir2019knowledge}
\bibfield{author}{\bibinfo{person}{Jamal~Abdul Nasir}, \bibinfo{person}{Iraklis
  Varlamis}, {and} \bibinfo{person}{Samreen Ishfaq}.}
  \bibinfo{year}{2019}\natexlab{}.
\newblock \showarticletitle{A knowledge-based semantic framework for query
  expansion}.
\newblock \bibinfo{journal}{\emph{Information Processing \& Management}}
  \bibinfo{volume}{56}, \bibinfo{number}{5} (\bibinfo{year}{2019}),
  \bibinfo{pages}{1605--1617}.
\newblock


\bibitem[\protect\citeauthoryear{Nunes, Ribeiro, and David}{Nunes
  et~al\mbox{.}}{2008}]%
        {nunes2008use}
\bibfield{author}{\bibinfo{person}{S{\'e}rgio Nunes}, \bibinfo{person}{Cristina
  Ribeiro}, {and} \bibinfo{person}{Gabriel David}.}
  \bibinfo{year}{2008}\natexlab{}.
\newblock \showarticletitle{Use of temporal expressions in web search}. In
  \bibinfo{booktitle}{\emph{European Conference on Information Retrieval}}.
  Springer, \bibinfo{pages}{580--584}.
\newblock


\bibitem[\protect\citeauthoryear{Padaki, Dai, and Callan}{Padaki
  et~al\mbox{.}}{2020}]%
        {padaki2020rethinking}
\bibfield{author}{\bibinfo{person}{Ramith Padaki}, \bibinfo{person}{Zhuyun
  Dai}, {and} \bibinfo{person}{Jamie Callan}.} \bibinfo{year}{2020}\natexlab{}.
\newblock \showarticletitle{Rethinking Query Expansion for BERT Reranking}. In
  \bibinfo{booktitle}{\emph{Advances in Information Retrieval: 42nd European
  Conference on IR Research, ECIR 2020, Lisbon, Portugal, April 14--17, 2020,
  Proceedings, Part II 42}}. Springer, \bibinfo{pages}{297--304}.
\newblock


\bibitem[\protect\citeauthoryear{Pass, Chowdhury, and Torgeson}{Pass
  et~al\mbox{.}}{2006}]%
        {pass2006aoldata}
\bibfield{author}{\bibinfo{person}{Greg Pass}, \bibinfo{person}{Abdur
  Chowdhury}, {and} \bibinfo{person}{Cayley Torgeson}.}
  \bibinfo{year}{2006}\natexlab{}.
\newblock \showarticletitle{A picture of search}. In
  \bibinfo{booktitle}{\emph{Proceedings of the 1st international conference on
  Scalable information systems}}. \bibinfo{pages}{1--es}.
\newblock


\bibitem[\protect\citeauthoryear{Radinsky, Diaz, Dumais, Shokouhi, Dong, and
  Chang}{Radinsky et~al\mbox{.}}{2013}]%
        {radinsky2013temporal}
\bibfield{author}{\bibinfo{person}{Kira Radinsky}, \bibinfo{person}{Fernando
  Diaz}, \bibinfo{person}{Susan Dumais}, \bibinfo{person}{Milad Shokouhi},
  \bibinfo{person}{Anlei Dong}, {and} \bibinfo{person}{Yi Chang}.}
  \bibinfo{year}{2013}\natexlab{}.
\newblock \showarticletitle{Temporal web dynamics and its application to
  information retrieval}. In \bibinfo{booktitle}{\emph{Proceedings of the sixth
  ACM international conference on Web search and data mining}}.
  \bibinfo{pages}{781--782}.
\newblock


\bibitem[\protect\citeauthoryear{Rehurek and Sojka}{Rehurek and Sojka}{2010}]%
        {gensim}
\bibfield{author}{\bibinfo{person}{Radim Rehurek} {and} \bibinfo{person}{Petr
  Sojka}.} \bibinfo{year}{2010}\natexlab{}.
\newblock \showarticletitle{Software framework for topic modelling with large
  corpora}. In \bibinfo{booktitle}{\emph{In Proceedings of the LREC 2010
  Workshop on New Challenges for NLP Frameworks}}. Citeseer.
\newblock


\bibitem[\protect\citeauthoryear{Robertson, Walker, Jones, Hancock-Beaulieu,
  Gatford, et~al\mbox{.}}{Robertson et~al\mbox{.}}{1995}]%
        {robertson1995okapi}
\bibfield{author}{\bibinfo{person}{Stephen~E Robertson}, \bibinfo{person}{Steve
  Walker}, \bibinfo{person}{Susan Jones}, \bibinfo{person}{Micheline~M
  Hancock-Beaulieu}, \bibinfo{person}{Mike Gatford}, {et~al\mbox{.}}}
  \bibinfo{year}{1995}\natexlab{}.
\newblock \showarticletitle{Okapi at TREC-3}.
\newblock \bibinfo{journal}{\emph{Nist Special Publication Sp}}
  \bibinfo{volume}{109} (\bibinfo{year}{1995}), \bibinfo{pages}{109}.
\newblock


\bibitem[\protect\citeauthoryear{Rosin, Adar, and Radinsky}{Rosin
  et~al\mbox{.}}{2017}]%
        {rosin2017learning}
\bibfield{author}{\bibinfo{person}{Guy~D Rosin}, \bibinfo{person}{Eytan Adar},
  {and} \bibinfo{person}{Kira Radinsky}.} \bibinfo{year}{2017}\natexlab{}.
\newblock \showarticletitle{Learning Word Relatedness over Time}. In
  \bibinfo{booktitle}{\emph{Proceedings of the 2017 Conference on Empirical
  Methods in Natural Language Processing}}. \bibinfo{pages}{1168--1178}.
\newblock


\bibitem[\protect\citeauthoryear{Rosin and Radinsky}{Rosin and
  Radinsky}{2019}]%
        {rosin2019generating}
\bibfield{author}{\bibinfo{person}{Guy~D Rosin} {and} \bibinfo{person}{Kira
  Radinsky}.} \bibinfo{year}{2019}\natexlab{}.
\newblock \showarticletitle{Generating Timelines by Modeling Semantic Change}.
  In \bibinfo{booktitle}{\emph{Proceedings of the 23rd Conference on
  Computational Natural Language Learning (CoNLL)}}. \bibinfo{pages}{186--195}.
\newblock


\bibitem[\protect\citeauthoryear{Roy, Paul, Mitra, and Garain}{Roy
  et~al\mbox{.}}{2016}]%
        {roy2016qe}
\bibfield{author}{\bibinfo{person}{Dwaipayan Roy}, \bibinfo{person}{Debjyoti
  Paul}, \bibinfo{person}{Mandar Mitra}, {and} \bibinfo{person}{Utpal Garain}.}
  \bibinfo{year}{2016}\natexlab{}.
\newblock \showarticletitle{Using word embeddings for automatic query
  expansion}.
\newblock \bibinfo{journal}{\emph{arXiv preprint arXiv:1606.07608}}
  (\bibinfo{year}{2016}).
\newblock


\bibitem[\protect\citeauthoryear{Setty and Hose}{Setty and Hose}{2018}]%
        {setty2018event2vec}
\bibfield{author}{\bibinfo{person}{Vinay Setty} {and} \bibinfo{person}{Katja
  Hose}.} \bibinfo{year}{2018}\natexlab{}.
\newblock \showarticletitle{Event2Vec: Neural embeddings for news events}. In
  \bibinfo{booktitle}{\emph{The 41st International ACM SIGIR Conference on
  Research \& Development in Information Retrieval}}.
  \bibinfo{pages}{1013--1016}.
\newblock


\bibitem[\protect\citeauthoryear{Shokouhi and Radinsky}{Shokouhi and
  Radinsky}{2012}]%
        {shokouhi2012time}
\bibfield{author}{\bibinfo{person}{Milad Shokouhi} {and} \bibinfo{person}{Kira
  Radinsky}.} \bibinfo{year}{2012}\natexlab{}.
\newblock \showarticletitle{Time-sensitive query auto-completion}. In
  \bibinfo{booktitle}{\emph{Proceedings of the 35th international ACM SIGIR
  conference on Research and development in information retrieval}}.
  \bibinfo{pages}{601--610}.
\newblock


\bibitem[\protect\citeauthoryear{Yamada, Asai, Sakuma, Shindo, Takeda,
  Takefuji, and Matsumoto}{Yamada et~al\mbox{.}}{2020}]%
        {wikipedia2vec}
\bibfield{author}{\bibinfo{person}{Ikuya Yamada}, \bibinfo{person}{Akari Asai},
  \bibinfo{person}{Jin Sakuma}, \bibinfo{person}{Hiroyuki Shindo},
  \bibinfo{person}{Hideaki Takeda}, \bibinfo{person}{Yoshiyasu Takefuji}, {and}
  \bibinfo{person}{Yuji Matsumoto}.} \bibinfo{year}{2020}\natexlab{}.
\newblock \showarticletitle{Wikipedia2Vec: An Efficient Toolkit for Learning
  and Visualizing the Embeddings of Words and Entities from Wikipedia}.
\newblock \bibinfo{journal}{\emph{arXiv preprint 1812.06280v3}}
  (\bibinfo{year}{2020}).
\newblock


\bibitem[\protect\citeauthoryear{Yamada, Shindo, Takeda, and Takefuji}{Yamada
  et~al\mbox{.}}{2016}]%
        {yamada2016joint}
\bibfield{author}{\bibinfo{person}{Ikuya Yamada}, \bibinfo{person}{Hiroyuki
  Shindo}, \bibinfo{person}{Hideaki Takeda}, {and} \bibinfo{person}{Yoshiyasu
  Takefuji}.} \bibinfo{year}{2016}\natexlab{}.
\newblock \showarticletitle{Joint Learning of the Embedding of Words and
  Entities for Named Entity Disambiguation}. In
  \bibinfo{booktitle}{\emph{Proceedings of The 20th SIGNLL Conference on
  Computational Natural Language Learning}}. \bibinfo{publisher}{Association
  for Computational Linguistics}, \bibinfo{pages}{250--259}.
\newblock
\urldef\tempurl%
\url{https://doi.org/10.18653/v1/K16-1025}
\showDOI{\tempurl}


\bibitem[\protect\citeauthoryear{Yao, Sun, Ding, Rao, and Xiong}{Yao
  et~al\mbox{.}}{2018}]%
        {yao2018dynamic}
\bibfield{author}{\bibinfo{person}{Zijun Yao}, \bibinfo{person}{Yifan Sun},
  \bibinfo{person}{Weicong Ding}, \bibinfo{person}{Nikhil Rao}, {and}
  \bibinfo{person}{Hui Xiong}.} \bibinfo{year}{2018}\natexlab{}.
\newblock \showarticletitle{Dynamic word embeddings for evolving semantic
  discovery}. In \bibinfo{booktitle}{\emph{Proceedings of the eleventh acm
  international conference on web search and data mining}}.
  \bibinfo{pages}{673--681}.
\newblock


\bibitem[\protect\citeauthoryear{Zamani and Croft}{Zamani and Croft}{2016a}]%
        {zamani2016embedding}
\bibfield{author}{\bibinfo{person}{Hamed Zamani} {and} \bibinfo{person}{W~Bruce
  Croft}.} \bibinfo{year}{2016}\natexlab{a}.
\newblock \showarticletitle{Embedding-based query language models}. In
  \bibinfo{booktitle}{\emph{Proceedings of the 2016 ACM international
  conference on the theory of information retrieval}}.
  \bibinfo{pages}{147--156}.
\newblock


\bibitem[\protect\citeauthoryear{Zamani and Croft}{Zamani and Croft}{2016b}]%
        {zamani2016estimating}
\bibfield{author}{\bibinfo{person}{Hamed Zamani} {and} \bibinfo{person}{W~Bruce
  Croft}.} \bibinfo{year}{2016}\natexlab{b}.
\newblock \showarticletitle{Estimating embedding vectors for queries}. In
  \bibinfo{booktitle}{\emph{Proceedings of the 2016 ACM International
  Conference on the Theory of Information Retrieval}}.
  \bibinfo{pages}{123--132}.
\newblock


\bibitem[\protect\citeauthoryear{Zamani and Croft}{Zamani and Croft}{2017}]%
        {zamani2017relevance}
\bibfield{author}{\bibinfo{person}{Hamed Zamani} {and} \bibinfo{person}{W~Bruce
  Croft}.} \bibinfo{year}{2017}\natexlab{}.
\newblock \showarticletitle{Relevance-based word embedding}. In
  \bibinfo{booktitle}{\emph{Proceedings of the 40th International ACM SIGIR
  Conference on Research and Development in Information Retrieval}}.
  \bibinfo{pages}{505--514}.
\newblock


\bibitem[\protect\citeauthoryear{Zhai and Lafferty}{Zhai and Lafferty}{2017}]%
        {zhai2017study}
\bibfield{author}{\bibinfo{person}{Chengxiang Zhai} {and} \bibinfo{person}{John
  Lafferty}.} \bibinfo{year}{2017}\natexlab{}.
\newblock \showarticletitle{A study of smoothing methods for language models
  applied to ad hoc information retrieval}. In \bibinfo{booktitle}{\emph{ACM
  SIGIR Forum}}, Vol.~\bibinfo{volume}{51}. ACM New York, NY, USA,
  \bibinfo{pages}{268--276}.
\newblock


\bibitem[\protect\citeauthoryear{Zhang, Han, and Lu}{Zhang
  et~al\mbox{.}}{2018}]%
        {zhang2018automatic}
\bibfield{author}{\bibinfo{person}{Xiaojuan Zhang}, \bibinfo{person}{Shuguang
  Han}, {and} \bibinfo{person}{Wei Lu}.} \bibinfo{year}{2018}\natexlab{}.
\newblock \showarticletitle{Automatic prediction of news intent for search
  queries}.
\newblock \bibinfo{journal}{\emph{The Electronic Library}}
  (\bibinfo{year}{2018}).
\newblock


\end{thebibliography}

\appendix
\section{Comparison to Pseudo-Relevance Feedback}
\neww{
Adding pseudo-relevance feedback (PRF) to an existing word embedding based QE method has been found to bring significant improvements to retrieval performance~\citep{kuzi2016query,zamani2016embedding}, and is thus consistently used by state-of-the-art methods~\citep{padaki2020rethinking,imani2019deep}. 
In this work, we presented a general model that is independent of PRF. 
We leave its integration with PRF for future work (i.e., PRF can be considered as an additional component on top of our model).
}

\neww{
Nevertheless, to extend our analysis, in this appendix we compare our method \tempo (which does not involve PRF) with a PRF method called Bo1, which is the default query expansion method in Terrier. Bo1 is a generalization of Rocchio’s method and is based on Divergence from Randomness~\citep{amati2002probabilistic}. Note that PRF requires an additional retrieval round and has access to more information about the query.
\Tempo achieves better or equal performance compared to Bo1 on all the TREC datasets used in this work, combined (though the differences are not statistically significant). On the P@10, NDCG, and MAP metrics, \tempo achieves 0.47, 0.48, and 0.31, compared to 0.42, 0.42, and 0.31 by Bo1, respectively.
}


\typeout{get arXiv to do 4 passes: Label(s) may have changed. Rerun}
\end{document}